\documentclass[fleqn,usenatbib]{mnras}

\usepackage{newtxtext,newtxmath}
\usepackage[T1]{fontenc}

\DeclareRobustCommand{\VAN}[3]{#2}
\let\VANthebibliography\thebibliography
\def\thebibliography{\DeclareRobustCommand{\VAN}[3]{##3}\VANthebibliography}

\usepackage{graphicx}	% Including figure files
\usepackage{amsmath}	% Advanced maths commands
\usepackage{mathtools}
\usepackage{gensymb}
\usepackage{CJKutf8}
\usepackage{longtable,booktabs,etoolbox,multirow,tabularx}
\title[Spectropolarimetry of SN 2019\lowercase{ein}]{Spectropolarimetry of the Type I\lowercase{a} 
{SN 2019\lowercase{ein}} rules out significant global asphericity of the ejecta} %of $^{56}\text{Ni}$}

\author[Patra et al.]{
Kishore C. Patra$^{1, \dagger,}$\thanks{E-mail: kcpatra@berkeley.edu},
Yi Yang\begin{CJK*}{UTF8}{gbsn}
(杨轶)
\end{CJK*}$^{1,2,\mathsection, \thanks{E-mail: yi.yang@berkeley.edu}}$,
Thomas G. Brink$^{1}$,
Peter H\"{o}flich$^{3}$,
Lifan Wang$^{4}$, \newauthor
Alexei V. Filippenko$^{1,5, \ddagger}$,   
Daniel Kasen$^{1,6}$,
Dietrich Baade$^{7}$,
Ryan J. Foley$^{8}$,
Justyn R. Maund$^{9}$, \newauthor
WeiKang Zheng$^{1}$, 
Tiara Hung$^{8}$, 
Aleksandar Cikota$^{10}$,
J. Craig Wheeler$^{11}$, 
Mattia Bulla$^{12}$
\\
$^{1}$Department of Astronomy, University of California, Berkeley, CA 94720-3411, USA\\
$^{2}$Department of Particle Physics and Astrophysics, Weizmann Institute of Science, Rehovot 76100, Israel\\
$^{3}$Department of Physics, Florida State University, 77 Chieftan Way, Tallahassee, FL 32306, USA\\
$^{4}$George P. and Cynthia Woods Mitchell Institute for Fundamental Physics $\&$ Astronomy, Texas A\&M University, 4242 TAMU, College Station, TX 77843, USA\\
$^{5}$Miller Institute for Basic Research in Science, University of California, Berkeley, CA 94720, USA\\
$^{6}$Departments of Physics and Astronomy, University of California, Berkeley; and Lawrence Berkeley National Laboratory, USA\\
$^{7}$European Organisation for Astronomical Research in the Southern Hemisphere (ESO), Karl-Schwarzschild-Str. 2, 85748 Garching b.\ M{\"u}nchen, Germany\\
$^{8}$Department of Astronomy and Astrophysics, University of California, Santa Cruz, CA 95064, USA\\
$^{9}$Department of Physics and Astronomy, University of Sheffield, Hicks Building, Hounsfield Road, Sheffield S3 7RH, UK \\
$^{10}$European Southern Observatory, Alonso de C{\'o}rdova 3107, Vitacura, Casilla 19001, Santiago de Chile, Chile \\
$^{11}$Department of Astronomy, University of Texas, Austin, TX 78712, USA \\
$^{12}$The Oskar Klein Centre, Department of Astronomy, Stockholm University, AlbaNova, SE-10691 Stockholm, Sweden \\
$^{\dagger}$Nagaraj-Noll Graduate Fellow \\
$^{\mathsection}$Bengier-Winslow-Robertson Fellow \\
$^{\ddagger}$Miller Senior Fellow\\
}

\date{Accepted XXX. Received YYY; in original form ZZZ}

\pubyear{2021}

\begin{document}
\label{firstpage}
\pagerange{\pageref{firstpage}--\pageref{lastpage}}
\maketitle

\begin{abstract}
Detailed spectropolarimetric studies may hold the key to probing the explosion mechanisms and the progenitor scenarios of Type Ia supernovae (SNe\,Ia). We present multi-epoch spectropolarimetry and imaging polarimetry of SN\,2019ein, an SN\,Ia showing high expansion velocities at early phases. The spectropolarimetry sequence spans from $\sim -11$ to $+$10\,days relative to peak brightness in the $B$-band. We find that the level of the continuum polarization of SN\,2019ein, after subtracting estimated interstellar polarization, is in the range 0.0--0.3\%, typical for SNe\,Ia. The polarization position angle remains roughly constant before and after the SN light-curve peak, implying that the inner regions share the same axisymmetry as the outer layers. We observe high polarization ($\sim 1\%$) across both the \ion{Si}{ii}\,$\lambda6355$ and \ion{Ca}{ii}\,near-infrared triplet features. These two lines also display complex polarization modulations. The spectropolarimetric properties of SN\,2019ein rule out a significant departure from spherical symmetry of the ejecta for up to a month after the explosion. These observations disfavour merger-induced and double-detonation models for SN\,2019ein. The imaging polarimetry shows weak evidence for a modest increase in polarization after $\sim 20$\,days since the $B$-band maximum. If this rise is real and is observed in other SNe\,Ia at similar phases, we may have seen, for the first time, an aspherical interior similar to what has been previously observed for SNe\,IIP. Future polarization observations of SNe\,Ia extending to post-peak epochs will help to examine the inner structure of the explosion.

\end{abstract}


%%%%%%%%%%%%%%%%%%%%%%%%%%%%%%%%%%%%%%%%%%%%%%%%%%

%%%%%%%%%%%%%%%%% BODY OF PAPER %%%%%%%%%%%%%%%%%%

\section{Introduction}
\label{sec:introduction}

During the last half century, Type Ia supernovae (SNe\,Ia; see \citealp{Filippenko_1997} and \citealp{Gal-Yam_2017} for reviews of SNe) have answered (and posed) a myriad of interesting questions in astrophysics. These range from nucleosynthesis, chemical enrichment \citep{Renzini_1999}, and heating of the interstellar medium \citep{Ciotti_etal_1991} to the discovery of the accelerating expansion of the universe \citep{Riess_etal_1998, Perlmutter_etal_1999}, and more recently the so-called ``Hubble tension" (as summarised by \citealp{Riess_2020_natrev}). Yet, the nature of the progenitor systems of SNe\,Ia is still unclear. It has been generally established that the rise of SNe\,Ia is powered by the thermonuclear runaway of white dwarfs (WDs; see \citealp{Hoyle_etal_1960, Howell_2011, Hillebrandt_etal_2013, Maoz_etal_2014, Hoeflich_2017, Soker_2019_review} for recent reviews). However, the exact mechanism by which a WD's explosion is triggered and propagates through the progenitor still remains poorly understood \citep{Arnett_1969, Nomoto_etal_1976, Khokhlov_etal_1991, Niemeyer_etal_1996, Reinecke_etal_2002, Plewa_etal_2004, Ropke_2007, Pakmor_etal_2011, Seitenzahl_etal_2013}. 

Multiple channels of progenitors have been theorised, including the double-degenerate scenario in which two WDs merge \citep{Iben_Tutukov_1984, Webbink_1984}, the single-degenerate scenario in which a WD accretes matter from a nondegenerate companion until the Chandrasekhar mass ($M_{\text{Ch}} \approx 1.4 ~M_{\odot}$) is approached \citep{Whelan_Iben_1973}, and tidal disruption of a WD by a compact companion and subsequent detonation of the WD \citep{Rosswog_etal_2009, Luminet_Pichon_1989}. 

Among these progenitor scenarios, a range of explosion mechanisms might be possible: delayed detonation, in which an initial deflagration front transitions to a detonation in a WD \citep{Khokhlov_etal_1991}; double detonation, where a thin He layer on the WD detonates first, starting a detonation front in the WD (see, for example, \citealp{Fink_etal_2010, Taam_1980, Shen_etal_2010}); and compressional heating of WDs triggered by the dynamic merger of two C-O WDs \citep{Pakmor_etal_2010, Hayden_etal_2010_shock} or head-on collisions of WDs \citep{Kushnir_etal_2013}. 

The shape of the ejecta and their circumstellar configuration is spatially unresolvable for extragalactic SN explosions, even with the best of ground-based interferometers\footnote{The minimum resolution required to study a nearby SN, for instance 3 Mpc away with a photosphere 100 au wide, would be $\sim10~ \mu$as. For comparison, the Event Horizon Telescope can achieve a resolution of $\sim60 ~\mu$as.}. Conventional photometry and spectroscopy provide a way to probe the kinematics and chemical structures of SN ejecta and their interaction with any pre-explosion circumstellar matter (CSM; see, for example, \citealp{Nugent_etal_2011, Gal-Yam_etal_2014}). However, these observations only offer crude clues on the structures of the ejecta and the interaction between the ejecta and any existing CSM. Such information is projected and smeared into the single dimension of radial velocity. Fortunately, spectropolarimetry, which measures polarization as a function of wavelength, provides a unique approach to the study of the SN explosion geometry. Any asphericity of the SN ejecta and the distribution of various elements formed in the ejecta are traced by the level of the continuum and the profiles of associated spectral lines in the polarization spectra, respectively. Additionally, the footprint of the interaction between the SN ejecta with any companion and CSM is encoded in the polarization spectra since such processes may create non-spherically- symmetric emission and/or scattering photon sources.

In SN atmospheres, photons are scattered by free electrons (Thomson scattering). The polarization state of the emitted photons is determined when they escape the last-scattering surface, known as the photosphere. A photon that undergoes Thomson scattering will be polarized perpendicularly to the plane of scattering, which is defined as the plane containing the incident and scattered rays. 
For a spatially-unresolved source, the observed polarization is an integration of the photons' electric vectors (E-vectors) projected in the plane of the sky. If the projected photosphere is circularly symmetric, a complete cancellation of the E-vectors results in zero net polarization. However, if the projected photosphere deviates from circular symmetry, incomplete cancellation of the E-vectors would lead to nonzero polarization across the spectrum. Additionally, any clumps of high-opacity absorbing material present above the photosphere along the observer's line of sight may block parts of the underlying photosphere. Therefore, an incomplete cancellation of the E-vectors will occur across the corresponding spectral lines, further producing nonzero polarization at the extinguished wavelengths.

SN\,2019ein [$\alpha$(J2000) = 13:53:29.11, $\delta$(J2000) = +40:16:31.33] was discovered on 2019 May 1.47 \citep[UT dates are used throughout this paper;][]{Tonry_etal_2019} in the outskirts of the nearby galaxy NGC 5353. The host of SN\,2019ein is a lenticular galaxy (Hubble type S0). A redshift of $z=0.00775$ taken from the NASA/IPAC Extragalactic Database\footnote{\url{https://ned.ipac.caltech.edu}} was adopted in this study. The spectrum obtained by the Las Cumbres Observatory (LCO) Global SN project on 2021 May 2.3 (about two weeks before the $B$-band light-curve peak) shows a very high expansion velocity of $\sim 24,000$\,km\,s$^{-1}$ as inferred from the absorption minimum of the \ion{Si}{ii} $\lambda6355$ line. The \ion{Ca}{ii} near-infrared triplet (hereafter \ion{Ca}{ii} NIR3) and the \ion{O}{i} lines are blended, creating a broad absorption profile. Curiously, the entire spectrum was slightly blueshifted with respect to the host-galaxy redshift, with the emission peaks of \ion{Si}{ii}, \ion{Ca}{ii}, and \ion{S}{ii} exhibiting velocities $\sim 10,000$\,km\,s$^{-1}$ toward the observer. \citet{Pellegrino_etal_2020} suggested that the apparent blueshift may be caused by an asymmetric explosion resulting in enhanced abundance of material at high velocities or due to optical-depth effects in the photosphere, in which most of the flux comes from material moving along the SN-Earth line of sight. The rise time of SN\,2019ein was short ($15.37 \pm 0.55$\,days), 
after which the SN faded rapidly with a 15\,day post-peak $B$-band magnitude decline \citep{Phillips_1993} $\Delta m_{15}(B) = 1.36 \pm 0.02$\,mag \citep{Kawabata_etal_2020}.

These features put SN\,2019ein in the rare company of high-velocity SNe\,Ia like SNe\,2004dt and 2006X, for which spectropolarimetric data have been obtained \citep{Wang_etal_2006_04dt, Patat_etal_2009}. In this work, we present five epochs of spectropolarimetry of SN\,2019ein. We describe our observations in Section~\ref{sec:observations} and present our results along with the analysis in Section~\ref{sec:analysis&results}. We discuss the interpretations of the data in Section~\ref{sec:discussion} and provide a concluding summary in Section~\ref{sec:concluding_summary}.

\section{Observations and data reduction}
\label{sec:observations}
\subsection{Kast Spectropolarimetry}

Spectropolarimetry of SN\,2019ein was obtained using the polarimetry mode of the Kast double spectrograph on the Shane 3\,m telescope at Lick Observatory \citep{Miller_etal_1988}. In the polarimetry mode, the light beam incident on the spectrograph is passed through a rotatable, half-wave plate and then through a Wollaston Prism. The prism splits the incident light into two perpendicularly polarized components, named the ordinary and the extraordinary beams, which appear on the detector as two parallel traces. Only the red channel of Kast was used for spectropolarimetry. A GG455 order-sorting filter was in place, blocking all first-order light below $\sim$4600 \AA\ and all second-order light below $\sim$9800 \AA. The usable wavelength range of the setup was 4600--9000 \AA. Observations were made with the 300 lines\,mm$^{-1}$ grating and the $3''$-wide slit, resulting in a resolution of $\Delta \lambda \approx 18$\,\AA\ ($\sim 800$\,km\,s$^{-1}$) at the central wavelength $\sim 6800$\,\AA.

Flatfield and arc-lamp exposures were obtained at the beginning of the observation night. The flatfield spectra were produced by the reflection of the light from an incandescent lamp off of the inner surface of the dome. 

SN\,2019ein and polarization standard stars were observed each night. Four exposures were carried out at retarder-plate angles of $0^\circ$, $45^\circ$, $22.5^\circ$, and $67.5^\circ$. Multiple sets of polarimetry exposures were obtained for SN\,2019ein to achieve higher signal-to-noise ratios (S/N). Since all observations were carried out at small airmasses ($\lesssim 1.25$; see Table~\ref{tbl:times}), we aligned the slit to the position angle of 180$^{\circ}$ (north-south direction). Because Kast does not have an atmospheric dispersion compensator (ADC) to atone for the differential loss of blue light \citep{Filippenko_1982}, the following sanity check was carried out. For each night, we compared the Stokes parameters measured for different sets of spectropolarimetry with the values derived for the set obtained at the smallest airmass, typically 1.05. The Stokes parameters for each set were consistent within the associated uncertainties, suggesting a negligible effect on the polarization measurement from the loss of blue light.

 Our nightly observations of the unpolarized standard star HD\,110897 confirmed the low instrumental polarization of the Kast spectrograph (Sec.~\ref{calibrations}). The polarizance test to determine the instrumental response to 100\% linearly polarized light was done by observing the same unpolarized standard star through a polarizing filter. Each night we also conducted spectropolarimetry of two polarization standard stars chosen among HD\,154445, HD\,161056, and HD\,155528 to determine the accuracy of polarimetric measurements (see Sec.~\ref{calibrations} for more details). A polarization ``probe star" was also observed to estimate the Galactic interstellar polarization (see Sec.~\ref{isp}).

Extraction of the ordinary and the extraordinary beams was carried out following standard techniques for CCD processing and spectrum extraction within IRAF\footnote{IRAF is distributed by the National Optical Astronomy Observatories,
    which are operated by the Association of Universities for Research
    in Astronomy, Inc., under cooperative agreement with the National
    Science Foundation (NSF).}. The images were bias subtracted using an overscan region.  Cosmic ray hits on the detector were removed with \textit{L.A.Cosmic} \citep{van_Dokkum_etal_2001}. For each night, flatfield images were combined and normalised by a low-order spline function to fit the continuum before applying to the science images. Then we used the prescription of \citet{Horne_1986} to optimally extract each spectrum independently from the science images with apertures typically set to width at $\sim 1$--2\% of the maximum of the spectrum profile. The background apertures were usually 5--10 pixels wide and placed 2--3 times the value of the full width at half-maximum intensity (FWHM) away from the profile centre. 
    
    Wavelength calibration was conducted separately for the ordinary and extraordinary beams in each individual exposure (all four retarder-plate angles) using lamp exposures. Small wavelength adjustments determined from the night-sky lines in the object frames were also applied to fine tune the wavelength calibration. A typical root-mean-square (RMS) accuracy of $\sim 0.2$\,\AA\ was achieved. Flux calibration of the ordinary and extraordinary beams of the SN was applied using the corresponding beam of a flux standard star observed at a similar airmass. We fit splines to the continuum of the flux-standard spectrum to generate a ``sensitivity function" that maps CCD counts to the flux at each wavelength. This mapping function was then applied to the SN spectra. Correction for telluric absorption was carried out by interpolating over the atmospheric absorption regions of the flux-standard spectrum.

\begin{table*}
\caption{Journal of spectropolarimetric observations.} 
\begin{tabular}{ccccccc}
	\hline 
	\hline
	UT Date & MJD & Phase$^a$ & Airmass Range & Avg. Seeing & Wavelength Range & Exposure Time$^b$   \\ 
	(MM-DD-YYYY)&   & (days) & & (arcsec) & (\AA) & (s)   \\ 
	\hline 
 05-05-2019 & 58607.3 & $-10.9$&1.04 --> 1.03& 1.30 &4570--9000& 4$\times$1080 \\
 05-06-2019 & 58608.3 & $-9.9$& 1.04 --> 1.13& 1.21 &4570--9000& 4$\times$1080 \\
 05-12-2019 & 58614.3 & $-3.9$&1.01 --> 1.14 & 1.63 &4570--9000& 4$\times$1080\\
 05-13-2019 & 58615.4 & $-2.9$& 1.03 --> 1.13 & 1.29 &4570--9000& 4$\times$1080\\
 05-26-2019 & 58628.4 & $+$10.1& 1.09 --> 1.25 & 1.86 & 4570--9000& 4$\times$1080\\
	\hline 
\end{tabular}\\
%\flushleft
{$^a$}{Relative to $B$-band peak brightness at MJD 58618.2 \citep{Kawabata_etal_2020}.} \\
{$^b$}{Number of waveplate positions $\times$ exposure time at each position.}
\label{tbl:times}
\end{table*}

\subsection{RINGO3 Imaging Polarimetry}

In this work, we adopted the imaging polarimetry of SN\,2019ein from \citet{Maund_etal_2021}. Details of the RINGO3 observations and data reduction are provided by \citet{Maund_etal_2021}. In brief, they obtained the data using the Liverpool Telescope (LT) located on the Canary Island of La Palma, with the RINGO3 polarimeter \citep{Arnold_etal_2012}. Observations were carried out in three cameras optimised to integrate over the following wavelength ranges: $b$, 3500--6400\,\AA;  $g$, 6500--7600\,\AA; and $r$, 7700--10000\,\AA. Interstellar polarization was subtracted from the Stokes parameters measured by \citet{Maund_etal_2021} (see Sec.~\ref{timeseries} for details). The Stokes parameters, polarization, and its position angle are shown in Table \ref{tbl:polsummary}.

\section{Analysis and results}
\label{sec:analysis&results}

\subsection{Calculating the Stokes $q$ and $u$}

We express the normalised Stokes parameters as $q = Q/I$ and $u = U/I$, where $Q$ and $U$ are the differences in flux with E-vector oscillating in two perpendicular directions and $I$ is the total flux. $U$ measures polarization along angles that are rotated by 45$^{\circ}$ with respect to those measured by $Q$. 

We calculate $q$ and $u$ from two sets of spectra obtained with the waveplate at [0$^{\circ}$, 45$^{\circ}$], and [22$^{\circ}$.5, 67$^{\circ}$.5], respectively. From the ordinary ($o$) and the extraordinary ($e$) flux beams ($f$), $q$ can be expressed as
\begin{equation}
q_{o} = \frac{f_{o,0} - f_{o,45}}{f_{o,0} + f_{o,45}} ~~ \mbox{and} ~~ 
q_{e} = \frac{f_{e,0} - f_{e,45}}{f_{e,0} + f_{e,45}},
\end{equation}
respectively, which are then averaged. 
Similarly, we calculate $u$ using the exposures at the other set of waveplate positions. 
The observed polarization is then given by
\begin{equation}
p_{\text{obs}} = \sqrt{q^{2} + u^{2}} 
\end{equation} 
and the polarization position angle ($PA$) is
\begin{equation}
PA_{\text{obs}} = \frac{1}{2} \arctan \left( {\frac{u}{q}} \right).
\end{equation}

The polarization defined this way is positive-definite and biased toward higher polarization, especially in the low S/N regime. The final derived polarization was achieved after a debiasing procedure following \citet{Wang_etal_1997}:
\begin{equation}
p = \left ( p_{\text{obs}} - \frac{\sigma^{2}_{p}}{p_{\text{obs}}} \right ) \times h(p_{\text{obs}} - \sigma_{p})  ~ \mbox{and} ~~ PA = PA_{\text{obs}},
\end{equation}
where $\sigma_{p}$ denotes the 1$\sigma$ uncertainty in $p$ and $h$ is the Heaviside step function. The flux spectrum is calculated by averaging all the spectra of $o$-rays and $e$-rays used in deriving $q$ and $u$. Figures~\ref{fig:addr}--\ref{fig:adh} show the measured Stokes $q$, $u$, $p$, $PA$, and total flux at each epoch.

\subsection{Polarimetric Calibration}
\label{calibrations}
The nightly-measured Stokes $q$ and $u$ of the unpolarized standard star HD\,110897 are consistent with a level of $<0.05\%$, indicating a low instrumental polarization and a high stability of the Kast spectropolarimeter over time. In the polarizance test with the same standard star, we determined that the polarimetric response of the instrument is larger than 99.5\% across the entire wavelength range (4600--9000\,\AA) and therefore does not necessitate further correction. The polarization position angle of SN\,2019ein was corrected as follows:
\begin{equation*}
\begin{multlined}
 q_{\text{corr}} = p_{\text{obs}}  ~ q_{\text{obs}} ~ \cos2({PA_{\text{obs}} - PA_{i}}), \\
u_{\text{corr}} = p_{\text{obs}}  ~ u_{\text{obs}} ~ \cos2({PA_{\text{obs}} - PA_{i}}), 
\end{multlined}
\end{equation*}
where $PA_{i}$ is the position angle of the instrumental polarization determined from the polarizance test. 

The polarization and position angle measurements of the two high-polarization standards observed on each night were respectively found to be within $0.1 \%$ and 3$^{\circ}$ of the references \citep{Schmidt_etal_1992, Wolff_etal_1996}.

\subsection{Interstellar Polarization}
\label{isp}
Light passing through interstellar dust clouds is polarized through dichroic extinction by nonspherical paramagnetic dust grains present along the line of sight. The contribution to polarization by dust, namely the interstellar polarization (ISP), must be removed to determine the intrinsic polarization of the source. Although several approaches have been commonly used to estimate the ISP along the SN-Earth line of sight (see, for example, \citealp{Stevance_etal_2019, Yang_etal_2020}), the exact level of the ISP contribution to the observed polarization of SN\,2019ein is generally uncertain. 

\citet{Serkowski_etal_1975} showed that the Galactic ISP can be constrained to $p_{\text{ISP}} < 9 \times E(B-V)$ \%. The Milky Way colour excess $ E(B-V)_{\text{MW}}$ along the line of sight of SN\,2019ein is 0.011\,mag \citep{Schlafly_etal_2011}, constraining ISP\textsubscript{MW} to  < 0.1\%. This upper limit is commensurate with the measured polarization of an ISP ``probe star''\footnote{We observed the star \emph{Gaia DR2\,1497177392672672128}.} -- an intrinsically unpolarized star within 1\degree\ of SN\,2019ein that probes at least 150\,pc scale height of the Galactic interstellar medium (ISM). The Stokes $q$ and $u$ measured for the probe star were found to be $< 0.05$\% in the continuum wavelength range of SN\,2019ein, indicating low contribution from Galactic reddening. 

An upper limit of the ISP from the host galaxy of SN\,2019ein (NGC 5353) 
can be estimated based on the host reddening $E(B-V)_{\rm host}=0.09\pm0.02$\,mag \citep{Kawabata_etal_2020}. Accounting for both the Galactic and SN\,2019ein-host dust, we place an upper limit of $p_{\rm ISP} < 0.9$\% along the SN-Earth line of sight. Such a value of ISP is higher than the continuum polarization level seen in SN\,2019ein; thus, interstellar dust could potentially account for all of the continuum polarization of SN\,2019ein. The caveat, however, is that Serkowski's law may not be applicable to all galaxies because different dust properties could lead to different efficiencies for ISP \citep{Leonard_etal_2002_ispeff}. Therefore, we employ a more direct approach by following the method used by \citet{Yang_etal_2020} to estimate the ISP Stokes parameters $q_{\rm ISP}$ and $u_{\rm ISP}$. 

We consider the wavelength region 4800--5600\,\AA\ in the spectrum when the SN is near its peak brightness. This region is expected to be intrinsically depolarized owing to multiple overlapping Fe absorption features, which create a ``line blanketing" effect whose opacity dominates over electron scattering \citep{Howell_etal_2001, Maund_etal_2013}. We set the level of Stokes $q$ and $u$ to 0 within this wavelength range on day $-2.9$ (see Fig.~\ref{fig:adg}), giving us an estimate of $q_{\rm ISP} \approx -0.24\% $ and $u_{\rm ISP} \approx 0.19\%$. We note that these estimates are consistent with the upper limit derived above using Serkowski's rule. These ISP values were subtracted from the observed $q$ and $u$ on each night and the polarization and the position angle were subsequently recalculated (see Table~\ref{tbl:polsummary}). We note that owing to the relatively low level of expected ISP, only a wavelength-independent ISP estimation is presented.
We will discuss the ISP-corrected continuum and line polarization of SN\,2019ein in Sections~\ref{contpol} and ~\ref{linepol}, respectively.

\begin{table*}
\centering
\caption{Summary of polarimetry results. }
\begin{tabular}{ccc|cccc|cccc}
	\hline 
	\hline
	 MJD & Phase$^a$ & Instrument & $q$(\%) & $u$(\%) & $p$(\%) & $PA$(deg) & $q$(\%) & $u$(\%) & $p$(\%) & $PA$(deg)  \\ 
	 
	 & (d) &&& \multicolumn{2}{c}{(observed)} &&& \multicolumn{2}{c}{(ISP corrected)} & \\
	\hline 
58607.3 & $-10.9$ & Kast   &   $-0.23(07)$ &0.05(07) &0.21(07) & 83(9)&   0.01(07) &$-0.14(07)$ &0.10(07) &138(15)   \\
58608.3 & $-9.9$ & Kast   & $-0.18(05)$ & 0.09(05) & 0.19(05)& 77(7)&   0.06(05) & $-0.10(05)$ &0.10(05) &151(12)   \\
58612.9 & $-5.3$ & RINGO3: $b$  &  $-0.50(16)$ & 0.43(15) & 0.64(15) & 70(7) &   $-0.13(25)$ &$-0.04(24)$ &$<0.01$ & $99(>360)$  \\
58614.3 & $-3.9$ & Kast  &   $-0.22(02)$ & 0.22(02)&0.31(02) &68(2) &   0.02(02) &0.03(02) &0.02(02) & 210(18)  \\
58615.4 & $-2.9$ & Kast  &   $-0.23(04)$ & 0.23(04)& 0.32(04)& 68(3)&  0.01(04)  & 0.03(04)& 0.00(04) & 215(30)  \\
58620.0 &  1.8 & RINGO3: $b$  &  $-0.33(15)$ & 0.64(16) & 0.70(15) & 59(6) &    0.04(24)& 0.17(25) &$<0.01$ & $219(>360)$  \\
58622.9 &  4.7 & RINGO3: $b$  &  $-0.28(15)$ & 0.34(13) & 0.42(12) & 65(9) &   0.09(24) & $-0.13(22)$ &$<0.01$ & $152(>360)$  \\
58628.4 & 10.1 & Kast  &   $-0.07(06)$  & $-0.05(07)$ & 0.03(07)& 108(24) &  0.17(06)  & $-0.24(07)$ &0.28(07) & 153(7)  \\
58630.9 & 12.7 & RINGO3: $b$  &  $-0.15(24)$ & 1.14(26) & 1.12(25) & 49(6) &   0.22(33) &0.67(35) &0.54(34) & 215(17)  \\
58639.0 & 20.8 & RINGO3: $b$  &  $-0.47(23)$ & $-0.91(23)$ & 0.99(22) & 122(7) &  $-0.10(32)$& $-1.38(32)$ & 1.31(32)& 133(6)  \\
	\hline 
\end{tabular} \\
$^a$Relative to $B$-band peak brightness at MJD 58618.2 \citep{Kawabata_etal_2020}.
\label{tbl:polsummary}
\end{table*}

\begin{figure}
	%\centering
	\includegraphics[width=1.1\linewidth]{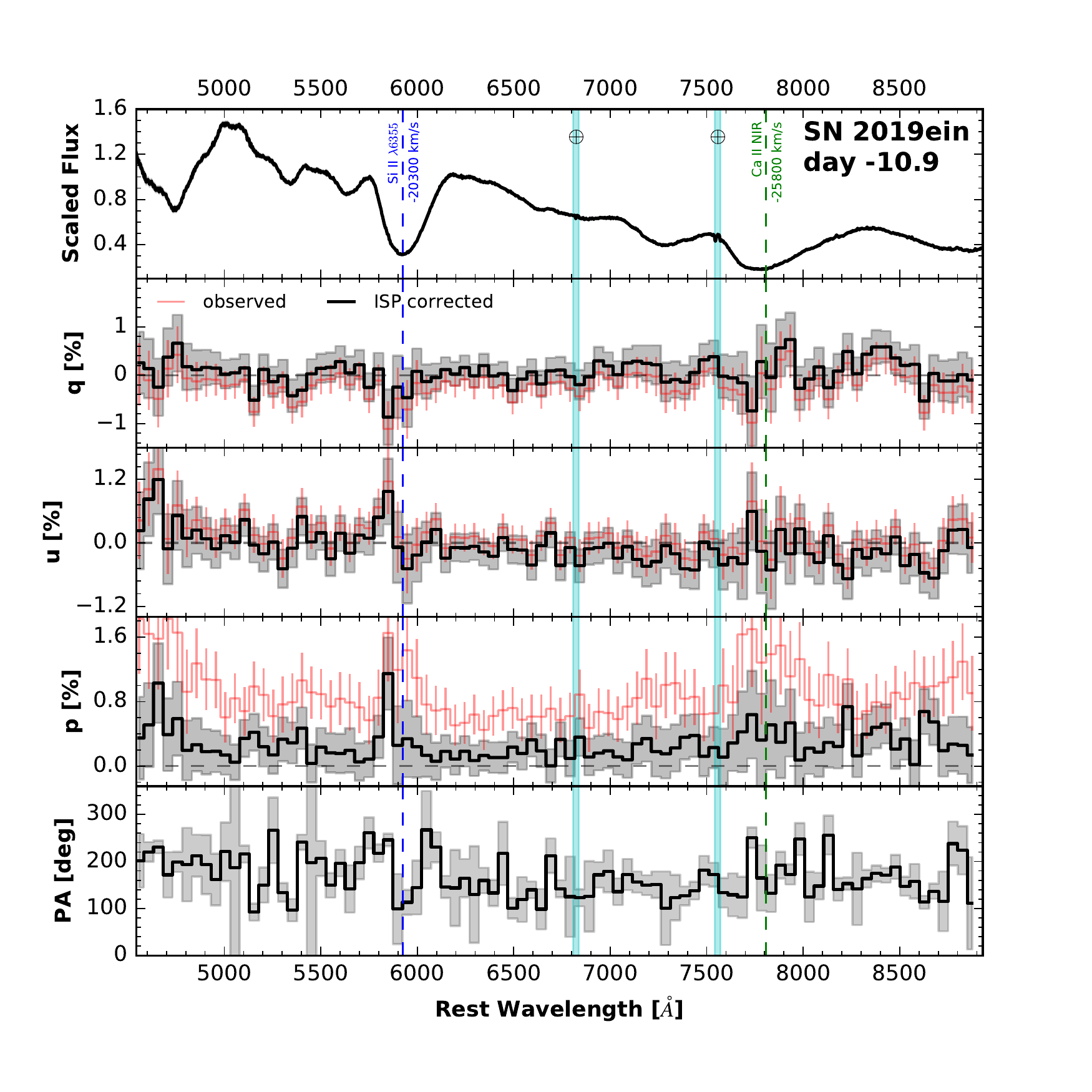}
	\caption{Spectropolarimetry of SN\,2019ein at day $-$10.9 relative to the $B$-band peak brightness at MJD 58618.2 \citep{Kawabata_etal_2020}. The cyan vertical bands represent the regions of telluric correction. The panels below the total-flux spectrum represent the polarimetry before (red) and after (black) the ISP correction. The grey-shaded area indicates the associated 1$\sigma$ uncertainty. The $PA$ panel shows only the polarization position angle after ISP correction. With the exception of the flux spectrum, we use 50\,\AA\ binning for the purpose of presentation.}
	\label{fig:addr}
\end{figure}

\begin{figure}
	\centering
	\includegraphics[width=1.1\linewidth]{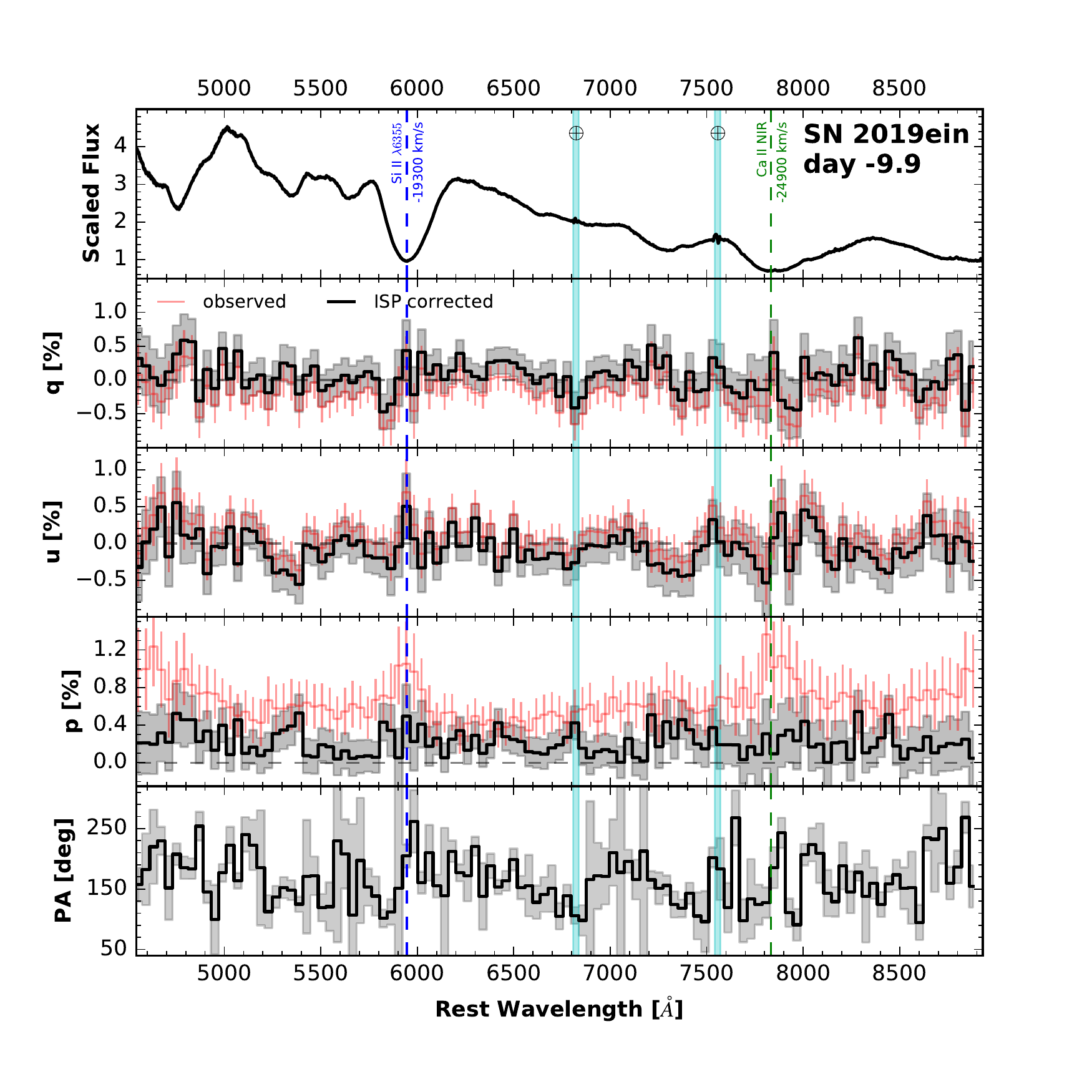}
	\caption{Similar to Figure \ref{fig:addr} but for day $-9.9$. We use 40\,\AA\ binning for the purpose of presentation.
	}
	\label{fig:addm}
\end{figure}

\begin{figure}
	\centering
	\includegraphics[width=1.1\linewidth]{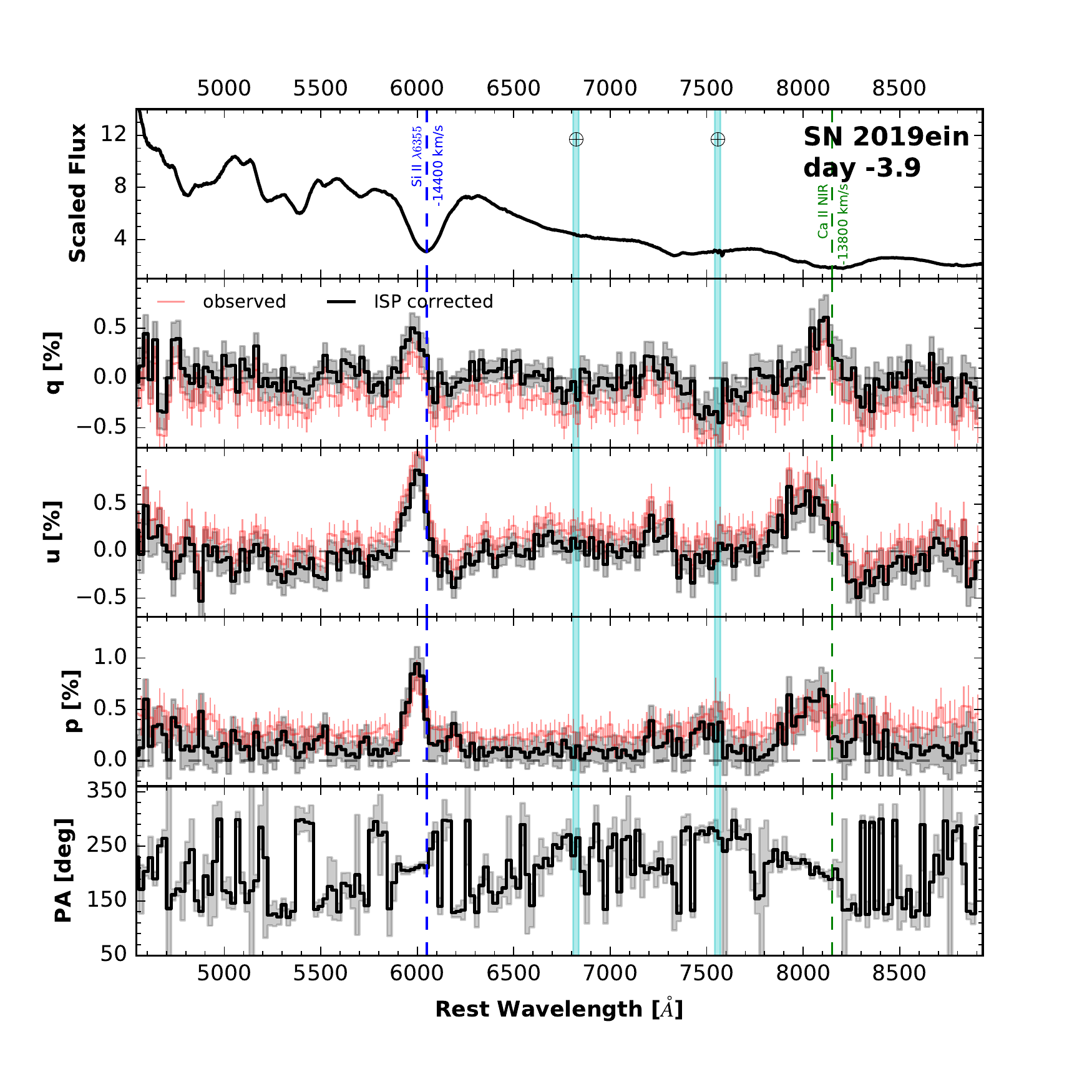}
	\caption{Similar to Fig.~\ref{fig:addr} but for day $-3.9$. We use 25\,\AA\ binning for the purpose of presentation.}
	\label{fig:adf}
\end{figure}

\begin{figure}
	\centering
	\includegraphics[width=1.1\linewidth]{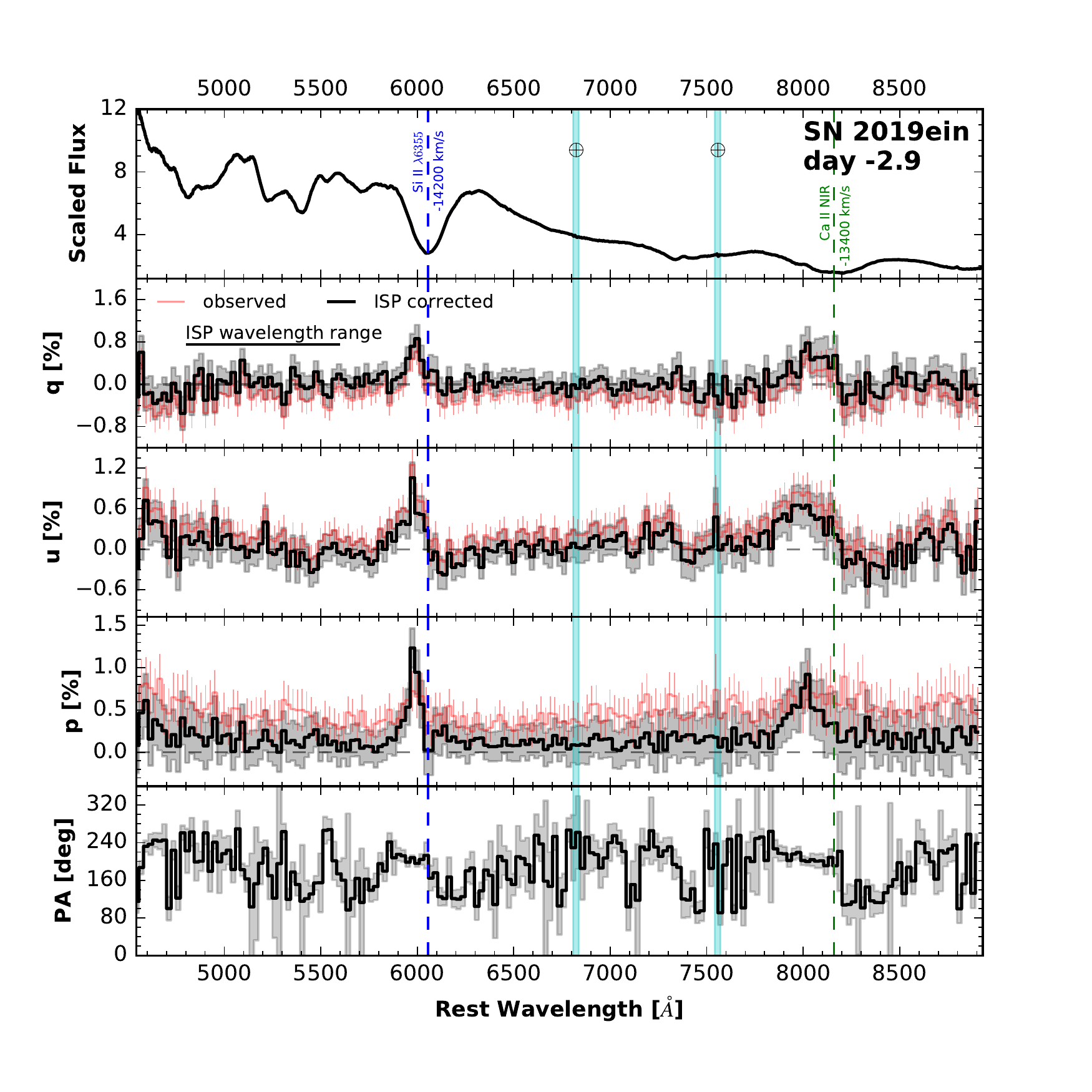}
	\caption{Similar to Fig.~\ref{fig:addr} but for day $-2.9$. We use 25\,\AA\ binning for the purpose of presentation. The wavelength region 4800--5600\,\AA\ was used to estimate the ISP as described in Section~\ref{isp}. }
	\label{fig:adg}
\end{figure}

\begin{figure}
	\centering
	\includegraphics[width=1.1\linewidth]{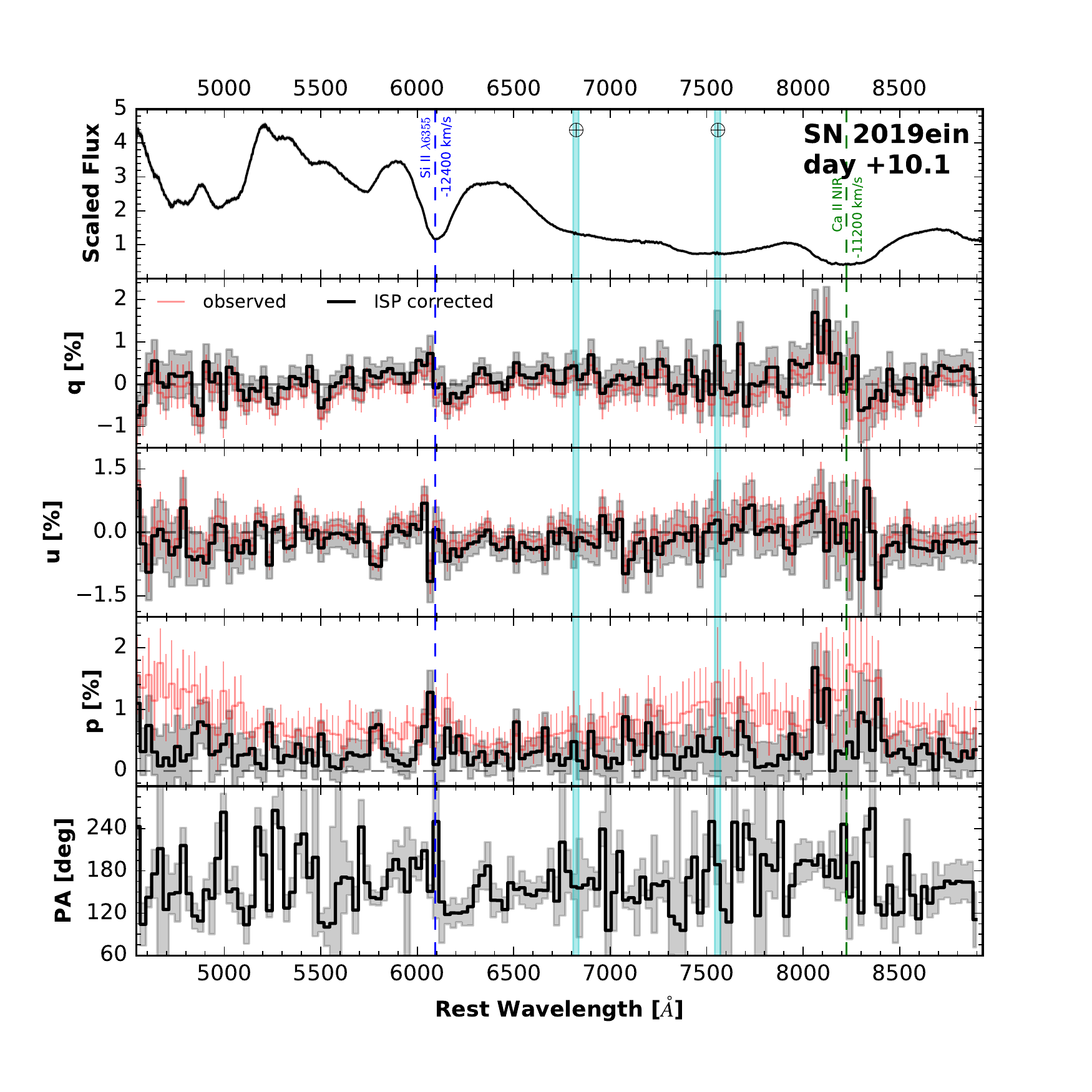}
	\caption{Similar to Fig.~\ref{fig:addr} but for day +10.1. We use 30\,\AA\ binning for the purpose of presentation.}
	\label{fig:adh}
\end{figure}

\subsection{Continuum Polarization}
\label{contpol}
Aspherical distribution of electrons, for instance an ellipsoidal photosphere, will cause imperfect cancellation of polarization E-vectors, leading to a nonzero continuum polarization \citep{Hoeflich_1991, Hoeflich_1995_94D, Bulla_etal_2015, Stevance_etal_2019}. SNe\,Ia typically show low continuum polarization ($\lesssim 0.3$\%; see, for example, \citealp{Wang_Wheeler_2008, Yang_etal_2020}), indicating that SNe\,Ia tend to be remarkably close to being spherical. 

The continuum polarization of SN\,2019ein and the associated uncertainty were estimated by binning the Stokes parameters over a wavelength range of 6400--7150\,\AA\ following a procedure similar to that described by \citet{Yang_etal_2020}. The selected spectral region is known to be free from strong absorption features \citep{Patat_etal_2009}. The uncertainty was correspondingly binned. 

The continuum polarization on days $-10.9$ and $-9.9$ is low with values of $p_{\text{cont}, -10.9\text{d}} = 0.10 \pm 0.07 \%$ and $p_{\text{cont}, -9.9\text{d}} = 0.10 \pm 0.05 \%$, respectively (Figs. \ref{fig:addr} and \ref{fig:addm}). The polarization is consistent with 0 as the SN approaches maximum brightness around day $-3.9$ and day $-2.9$ (Figs. \ref{fig:adf} and \ref{fig:adg}). After peak brightness, polarization increases slightly, reaching $p_{\text{cont}, +10.1\text{d}} = 0.28 \pm 0.07 \%$ (Fig. \ref{fig:adh}). 
We note that these values are consistent with infrared spectropolarimetry of SN\,2019ein, which found a 3$\sigma$ upper limit on polarization of 1.2\% around the SN peak brightness \citep{Tinyanont_2021_IR}.

The measured continuum polarization position angle ($PA$), which represents the position of global axisymmetry of the ejecta, remains fairly consistent before and after peak brightness, hovering around $150^{\circ}$. Even though the $PA$ on days $-3.9$ and $-2.9$ is apparently larger ($\sim 210^{\circ}$), we note that the polarization is so close to zero that the $PA$ is essentially undetermined on those days.

\subsection{Line Polarization}
\label{linepol}
Polarization signal at specific spectral lines arises due to the presence of clumps of material above the photosphere. The absorbing material partially obscures the underlying Thomson-scattering photosphere, resulting in an excess of polarization superimposed on the continuum polarization over the range of the absorption wavelengths. Below, we present the polarization of three absorption features as follows.

\begin{enumerate}
\item \ion{Si}{ii}\,$\lambda$6355: We observe weak line polarization ($0.5 \pm 0.4 \%$) on day $-$10.9, whereas no significant line polarization was detected on day $-$9.9. On days $-$3.9 and $-$2.9, we see strong line polarization, which reaches its peak value of $\sim 1\%$ at an expansion velocity of $\sim 17,000$\,km\,s$^{-1}$. The strong line polarization persists into day 10.1, at an expansion velocity of $\sim 13,600$\,km\,s$^{-1}$. 

\item \ion{O}{i}\,$\lambda 7774$: No significant line polarization was seen at any phase. 

\item \ion{Ca}{ii} NIR3: Similarly to \ion{Si}{ii}, no significant line polarization was observed on days $-$10.9 and $-$9.9. In contrast, on days $-$3.9 and $-$2.9, we see strong line polarization, reaching its peak value of $\sim 0.8\%$ at an expansion velocity of $\sim 18,000$\,km\,s$^{-1}$. The strong line polarization continues into day 10.1, reaching up to $\sim 1.5\%$ at 17,000\,km\,s$^{-1}$. 

\end{enumerate}

Starting on day $-3.9$ and thereafter, both \ion{Si}{ii} and \ion{Ca}{ii} show a complex structure in the polarization spectra, likely associated with high-velocity (HV) and normal-velocity (NV) components. 
For example, at day $+$10.1, the polarization across \ion{Ca}{ii} NIR3 reached two local maxima at $-$17,000\,km\,s$^{-1}$ and $-$5900\,km\,s$^{-1}$. The emergence of the line polarization of both the HV and NV components over time may result from an increase of \ion{Si}{ii} and \ion{Ca}{ii} opacity from larger to smaller radii.

\subsection{The $q$--$u$ Plane and the Dominant Axis}
\label{quplane}

\begin{table*}
\centering
\caption{Fitted parameters for the dominant axes.} 
\begin{tabular}{c|cc|ccc|ccc}

	\hline 
	\hline
	Phase &  $\alpha_{\text{cont}}$  & $\beta_{\text{cont}}$ & $\alpha_{\text{Si II}}$ & $\beta_{\text{Si II}}$ & Velocity Range  &  $\alpha_{\text{Ca II NIR3}}$ & $\beta_{\text{Ca II NIR3}}$ & Velocity Range  \\ 
	(days) & & & & & ($-10^{3}$\,km\,s$^{-1}$) & & & ($-10^{3}$\,km\,s$^{-1}$) \\
	\hline 
 $-10.9$  & $0.81(21)$ & $-0.15(04)$ & $-2.23(1.35)$ & $-0.12(23)$ & $28.6$ -- $9.7$   & $-0.74(24)$ & $-0.11(07)$ & $34.8$ -- $6.7$ \\
 
 $-9.9$  & 4.68(2.62) & $-0.47(24)$ & 0.99(49)& $-0.10(09)$ & $28.5$ -- $7.3$  & 0.87(28) & $-0.08(06)$ & $34.8$ -- $6.7$\\
 
 $-3.9$  & 0.33(09) & $-0.04(01)$ & 1.95(27)& $-0.06(05)$ & $23.9$ -- $5.0$  & 2.24(52) & $0.00(08)$ & $26.1$ -- $3.2$\\
 
 $-2.9$  & $-0.94(14)$ & 0.03(02) & 1.37(24)& $-0.09(06)$ &  $23.8$ -- $3.9$ & 1.33(23) & $-0.01(07)$ & $26.1$ -- $3.2$\\
 
 +10.1  & 0.02(13) & $-0.20(04)$ & 0.90(31)& $-0.23(08)$ & $21.5$ -- $2.6$  & 0.62(11) & $-0.28(06)$ & $22.5$ -- $0.3$\\
 
	\hline 
\end{tabular} 

\label{tbl:domaxis_params}
\end{table*}

\begin{figure*}
	\centering
	\includegraphics[width=0.8\linewidth]{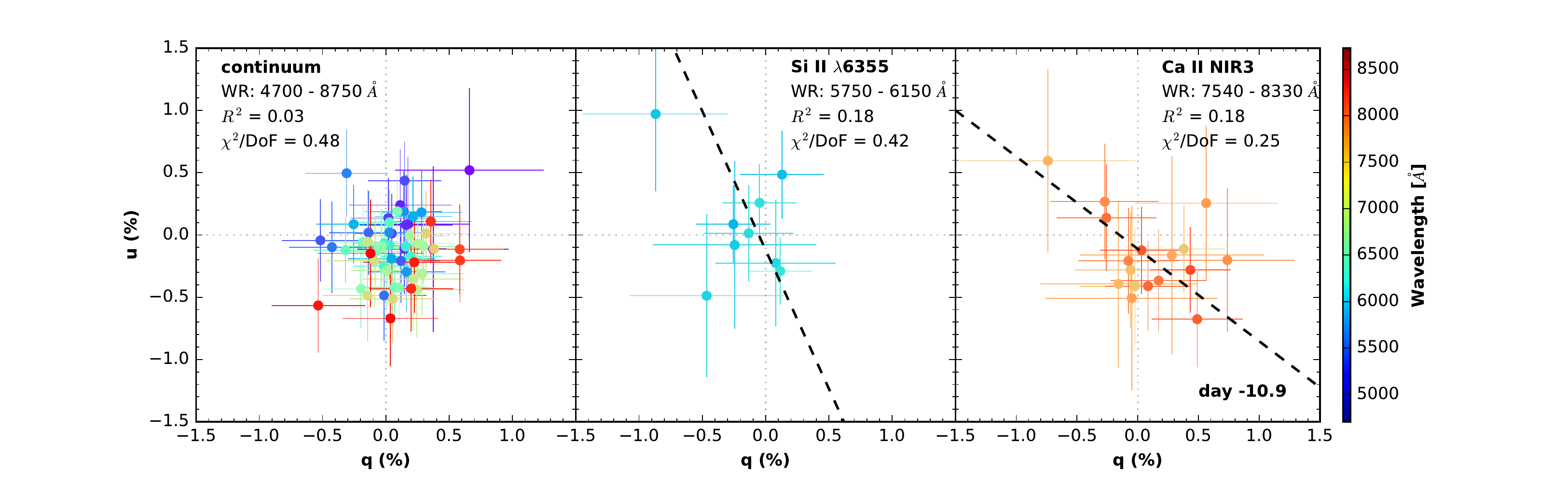}
	\includegraphics[width=0.8\linewidth]{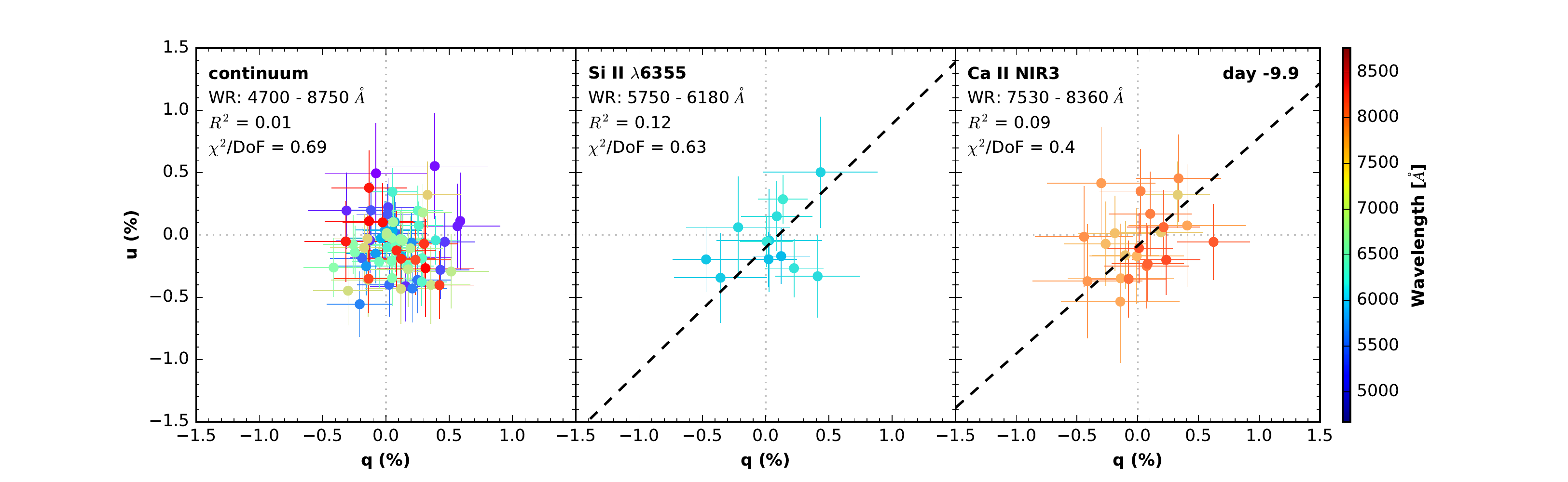}
	\includegraphics[width=0.8\linewidth]{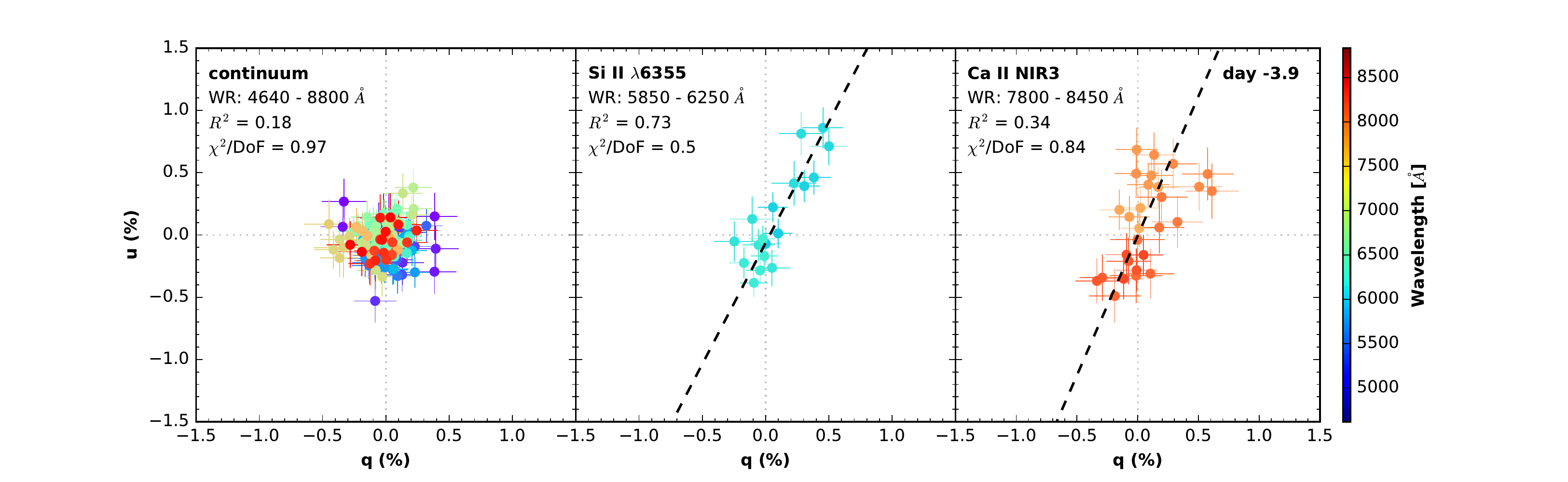}
	\includegraphics[width=0.8\linewidth]{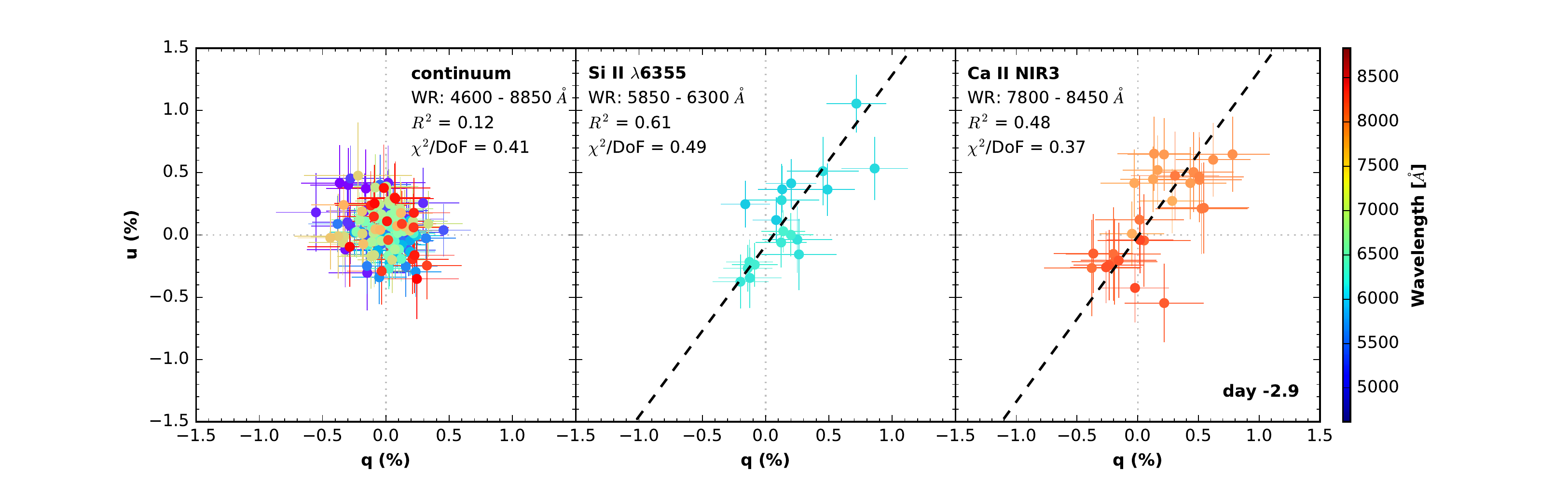}
	\includegraphics[width=0.8\linewidth]{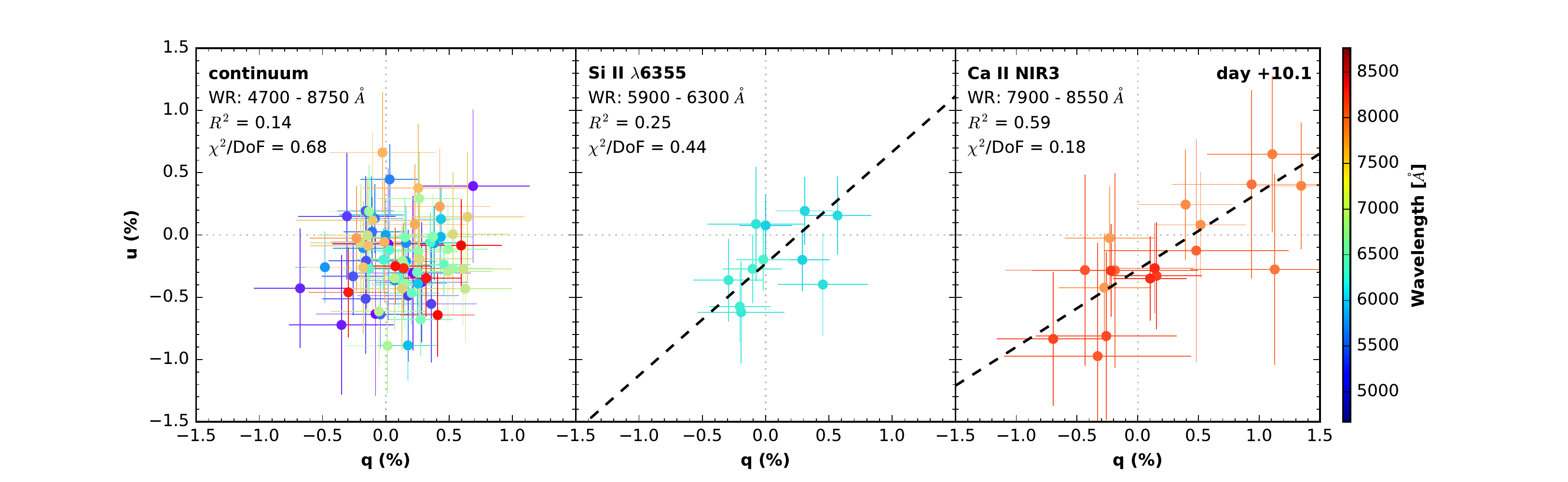}
	\caption{Polarization in the Stokes $q$--$u$ plane. The \ion{Si}{ii} $\lambda$6355 and \ion{Ca}{ii} NIR3 lines were omited from the left-side panels labeled ``continuum''. The dashed lines in the ``\ion{Si}{ii} $\lambda$6355'' and the ``\ion{Ca}{ii} NIR3'' panels represent the dominant axes for the labeled absorption features. }
	\label{fig:all_qu}
\end{figure*}

Plotting the Stokes parameters in the $q$--$u$ plane allows us to examine the axisymmetry of the continuum and various spectral features. If the SN ejecta are smooth and axisymmetric, the data points should fall along a straight line called the ``dominant axis" \citep{Wang_etal_2001, Wang_etal_2003_01el}. Deviations from the dominant axis in the perpendicular direction represent departures from axisymmetry and clumpiness of the ejecta. The dominant axis is determined by
\begin{equation}
    u = \alpha q + \beta,
\end{equation}
where $\alpha$ and $\beta$ are the fitted parameters from an error-weighted orthogonal distance regression.
In Figure~\ref{fig:all_qu}, we present the polarization in the $q$--$u$ plane in the continuum as well as for \ion{Si}{ii} $\lambda$6355 and \ion{Ca}{ii} NIR3. We omitted the \ion{Si}{ii} and \ion{Ca}{ii} lines when plotting the continuum $q$ and $u$ in the wavelength range 4700--8750 \AA. The fitted parameters $\alpha$ and $\beta$ that characterise the dominant axes are given in Table \ref{tbl:domaxis_params}.

As suggested by the values of $\chi^2/DoF$ in Figure~\ref{fig:all_qu}, the departure from the dominant axis fitted across the line profile indicates a significant clumpiness in the \ion{Si}{ii}-rich ejecta on days $-$10.9 and $-$9.9. Considering the relatively large values of $\chi^2/DoF$, it is ambiguous whether a dominant axis is present at early times. The absence of a clear dominant axis together with the measured low polarization suggests that any \ion{Si}{ii}-rich clumps are fairly uniformly distributed in the ejecta at early times, when the photosphere only intersects with the outermost part of the ejecta. Axial symmetry is evident on days $-$3.9 and $-$2.9 for \ion{Si}{ii} when the SN is near its peak brightness. However, the axial symmetry becomes weak and clumpiness increases again around day +10. \ion{Ca}{ii} shows an overall higher degree of clumpiness compared to \ion{Si}{ii}. \ion{Ca}{ii} also exhibits weak axial symmetry at early times. However, as time progresses, \ion{Ca}{ii} settles onto a more prominent symmetry axis. Unlike \ion{Si}{ii}, \ion{Ca}{ii} continues to exhibit high axial symmetry at day +10. The symmetry axes of both \ion{Si}{ii} and \ion{Ca}{ii} are roughly aligned with each other and remain fairly constant starting on day $-10$ and thereafter.

As seen in the left panels of Figure~\ref{fig:all_qu}, no clear dominant axis can be identified in the continuum at any epoch. The data points form a cloud centred near the origin. This strengthens the case that even though the overall ejecta geometry is spherical, the polarization in \ion{Si}{ii} and \ion{Ca}{ii} is due to clumps of explosively synthesised material.

\subsection{Polarization Time Series}
\label{timeseries}
\begin{figure*}
	\centering
	
	\includegraphics[width=0.8\linewidth]{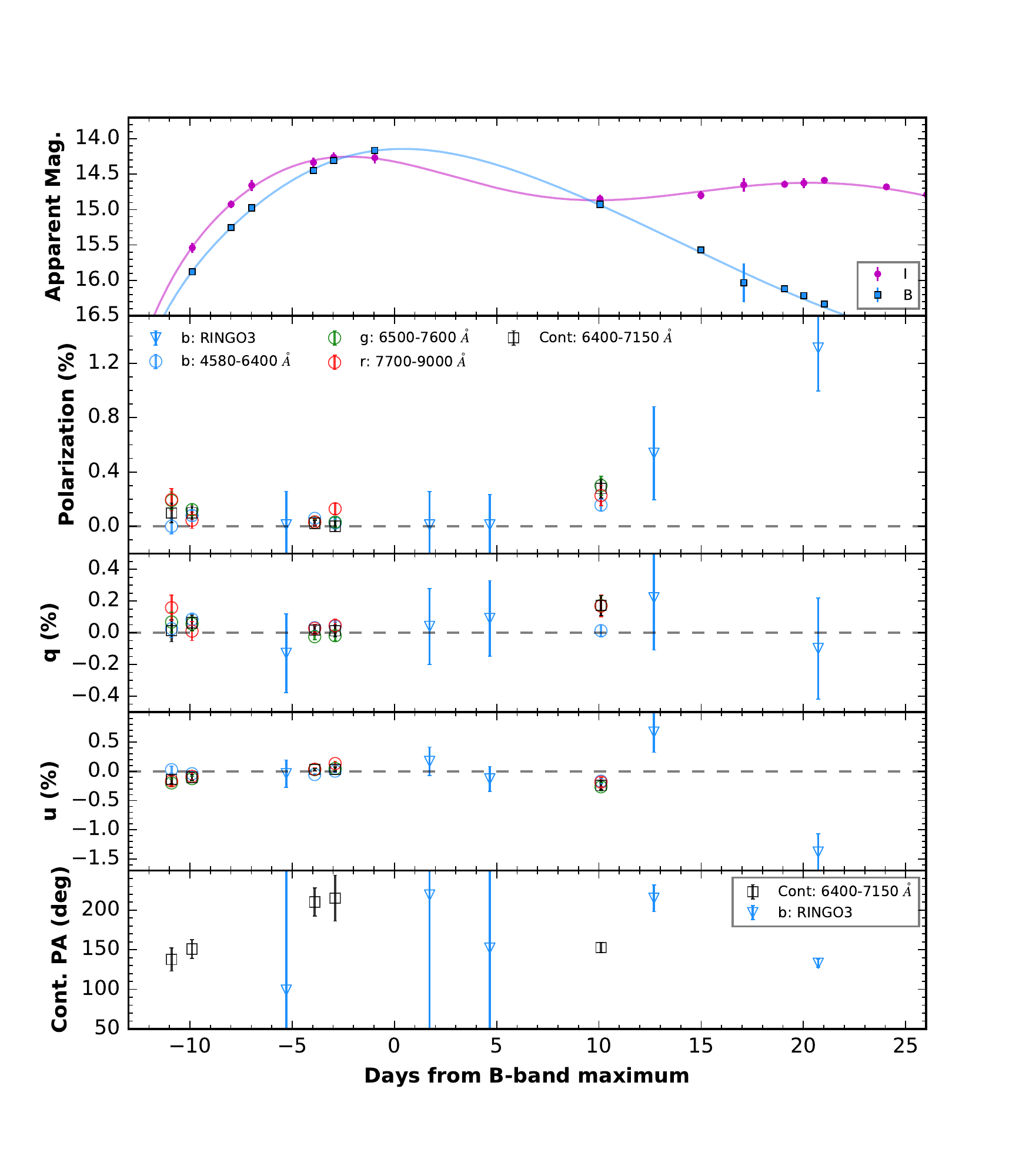}
	\caption{Polarization vs. time. 
	%The Kast spectropolarimetric measurements combined with RINGO3 imaging polarimetry \citep{Maund_etal_2021} show that the polarization of SN\,2019ein remains low until about day +15. The polarization appears to increase thereafter. 
	The polarization and $PA$ have been corrected for the ISP. The black squares represent the continuum polarization and $PA$ from the Kast spectropolarimeter. The red, blue, and green circles show the polarization measured by Kast in different wavelength regions: blue (4580--6400\,\AA), green (6500--7600\,\AA), and red (7700--9000\,\AA). The blue inverted triangles show polarization and $PA$ measured by RINGO3 in the blue channel with wavelength range 3500--6400\,\AA. The light curves in the top panel were obtained by KAIT at Lick Observatory.}
	\label{fig:synpol}
\end{figure*}

We build a temporal series of polarization measurements of SN\,2019ein by combining the Kast spectropolarimetry and the RINGO3 imaging polarimetry. In order to compare the polarization measured by the two instruments, we binned the Kast spectropolarimetry over the RINGO3 $b$, $g$, and $r$ filter passbands. This process estimates the equivalent imaging polarimetry data points in RINGO3 filters. In this way, we built the polarimetric dataset with a time baseline from $-$11 to 21 days relative to the $B$-band light-curve peak of SN\,2019ein. The broad-band polarization from Kast was derived by integrating over the wavelength of the filter-transmission-weighted polarized flux. 

The combined polarization time series is presented in Figure~\ref{fig:synpol}. The top panel displays two light curves obtained in the Landolt $I$ and $B$ bands with the Katzman Automatic Imaging Telescope (KAIT; \citealp{filippenko_etal_2001_kait}) at Lick Observatory. The middle panel shows polarization over time. The red, blue, and green circles represent the synthesised Kast polarization in wavelength ranges that roughly match the three channels of RINGO3. The black squares present the Kast continuum polarization in the wavelength range 6400--7150\,\AA. We note that the $b$ and $r$ bins include the polarization of the \ion{Si}{ii} $\lambda6355$ and \ion{Ca}{ii} NIR3 features, respectively. The blue inverted triangles show the polarization measured by the blue channel of RINGO3. The bottom panel provides the measured $PA$ in the continuum region (6400--7150\,\AA) by Kast and by the blue channel of RINGO3. The polarimetry presented here has been ISP corrected. Since we do not know the exact magnitude of any systematic bias (for example, instrumental polarization) in RINGO3 measurements, we employed a different strategy to account for ISP and any systematic bias: we calculated the mean $q$ and $u$ from the three epochs within $\sim 5$\,days of the $B$-band peak brightness (days $-5.3$, $+$1.8, and $+$4.7; see Table \ref{tbl:polsummary}), 
which gives $q_{\text{ISP+sys}} \approx -0.37\%$ and $u_{\text{ISP+sys}} \approx 0.47\%$. We then subtracted the averaged $q$ and $u$ from the observed Stokes parameters of all RINGO3 data under the assumption that SNe\,Ia exhibit effectively zero continuum polarization near peak brightness. This assumption is validated independently by the Kast spectropolarimetry of SN\,2019ein on days $-2.9$ and $-3.9$. We also propagated the uncertainty of ISP subtraction into the final calculations of $p$ and $PA$.

\section{Discussion}
\label{sec:discussion}

A nonzero continuum polarization may result from either an overall inhomogeneous electron density distribution or a nonspherical heating source. The latter case was seen in models of \citet{Bulla_etal_2016a, Bulla_etal_2016b} for SNe\,Ia, and has also been used to explain the observed increase in polarization during the plateau phase of SNe\,IIP, in which an aspherical ionisation front of $^{56}$Ni is typically present \citep{Hoeflich_etal_1996_93J}.

%For SNe\,Ia, a low level of continuum polarization persists for the period of SN brightening, suggesting a globally spherical ejecta distribution  (e.g., \citealp{Wang_Wheeler_2008}). 
Overall, 
the polarization properties of SN\,2019ein before its peak luminosity are 
consistent with the typical behaviour of SNe\,Ia. For example, the 
continuum polarization is $< 0.2$\%. Distinct line polarization, 
which is typically on the order of 1\%, can also be identified across some 
prominent spectral lines including 
\ion{Si}{ii}, \ion{Fe}{ii}, and \ion{Ca}{ii} (\citealp{Wang_Wheeler_2008}). 
The asymmetric distribution of the intermediate-mass elements 
(IMEs, 9$\leqslant$Z$\leqslant$20, including Si, Ca, S, and Mg)
inferred from the line polarization is indicative of sufficient outward mixing. IMEs generated in the nuclear burning can also be produced 
at higher velocities compared to the SN ejecta.
In thermonuclear SNe, below the production zone of the IMEs, the inner 
burning region is surrounded by C and O from the progenitor WD. 
The energy input is given by the radioactive decay chain of 
$^{56}$Ni$\rightarrow^{56}$Co$\rightarrow^{56}$Fe, which is initiated by 
the nucleosynthesis of $^{56}$Ni as the main product of the silicon-burning 
process. An asymmetric $^{56}$Ni distribution hence results in an 
asymmetric energy source. %The low continuum polarization observed in SN\,2019ein implies a spherical $^{56}$Ni distribution, sans any optical-depth effects. 

The low line and continuum polarization at early times (day $\sim -11$) do not support the idea put forward by \citet{Pellegrino_etal_2020} that the bluesfhited emission peaks in early-time spectra are due to an aspherical explosion enhancing abundance of material at high velocities. Furthermore, the lack of a clear dominant axis in the $q$--$u$ plane at early times indicates that even if clumps of high-velocity material are present, they must be fairly uniformly distributed in the outer ejecta. We cannot, however, rule out the possibility that the apparent blueshift is due to optical-depth effects arising from a steep density profile in the ejecta  \citep{Pellegrino_etal_2020}.

As seen in the bottom panel of Figure~\ref{fig:synpol}, we do not identify 
significant variation in $PA$ at different epochs. 
We note that near peak brightness, $q$ and $u$ intrinsic to SN\,2019ein are very close to zero after correcting for the ISP as discussed in Section~\ref{timeseries}. Such small values of Stokes parameters lead to effectively random values of the $PA$ around the SN light-curve peak.
Furthermore, we remark that the $PA$ calculated before ISP removal also tends to be consistent from day $-$11 to $+$21 (see Table \ref{tbl:polsummary}). A relatively low level of continuum polarization together with a roughly constant direction of the dominant axis suggests a common axial symmetry from the outer electron-scattering zone to the inner region near the energy source. %Significant departures from spherical symmetry can also be ruled out throughout the ejecta up until $\sim 20$\,days after maximum brightness. 

We remark that RINGO3 uses two dichroics to separate the 
three wavelength channels and a depolarizing Lyot prism, resulting in an 
induced systematic uncertainty in polarization of up to $\sim 0.5$\% 
\citep{Jermak_2017}. As shown in the third and the fourth panels of Figure~\ref{fig:synpol}, after day +5, the continuum level of polarization estimated from RINGO3 observations is mostly from $u$, while $q$ is consistent with zero. A moderate degree of asphericity is 
suggested by Kast spectropolarimetry at day $+$10. Unfortunately, we were not able to conduct Kast spectropolarimetry of SN\,2019ein after day $+$10 owing to technical issues. Therefore, the ``rise" in polarization after maximum brightness is anchored by just one RINGO3 measurement from day $+$21. For this reason, out of an abundance of caution, we refrain from claiming that a definitive rise in polarization was observed in SN\,2019ein post maximum brightness. However, if the post-peak rise of polarization is real and intrinsic to SN\,2019ein, we cautiously provide our interpretations below, hoping to invite more sophisticated theoretical investigations.

We suggest that the secondary maximum in the NIR luminosity could 
be the key for understanding the post-peak rise of the continuum polarization. 
The formation of the secondary maximum in the NIR can be understood as an 
opacity effect. To first order, the NIR luminosity is given by 
$L_{\text{NIR}}(t) \propto T(t)^4 \times R(t)^2$, where $T(t)$ and $R(t)$ represent 
the temperature and the radius of the photosphere as a function of time, 
respectively. In normal SNe\,Ia, although the photospheric radius increases 
with time because the opacity remains high for $\sim 2$--3 weeks after the 
SN explosion, it recedes gradually in mass coordinate (see, e.g., Fig.~3 of 
\citealp{Hoeflich_2017}) until the SN light curve approaches its secondary 
maximum in the NIR. This may explain the relatively little evolution of 
polarization seen in SN\,2019ein until peak brightness. 

After peak brightness, the recession of the SN photosphere is governed 
by the geometrical dilution of the homologously expanding envelope. However, 
$R(t)$ decreases rapidly owing to the significant drop in opacity when Fe-group 
elements begin recombining from ionisation states III to II 
\citep{Hoeflich_etal_1993, Hoeflich_etal_1998, Kasen_etal_2006}. Therefore, 
when the SN reaches its secondary NIR maximum $\sim 40$--50\,days 
after the explosion, most of the energy 
input emerges above the photosphere. In the presence of an asymmetric energy 
source, the flux at the photosphere will be direction-dependent. 
Consequently, an inhomogeneous photosphere could develop, leading to rise in polarization of the SN after maximum brightness. The interpretation of any rise in late-time polarization may be complicated by optical-depth effects that still remain poorly understood \citep{Hoeflich_1991} in SNe photospheres. As such, detailed theoretical investigations into how polarization behaves over time for various explosion mechanisms are pressingly warranted.

On the other hand, if the rise in late-time polarization is shown to be not true and the inner energy source is found to be spherical, we also arrive at an interesting implication. In that case, models with burning starting on the surface might be better at explaining the observed polarimetry. We speculate that if the burning starts on the surface, the detonation will propagate through the central zone supersonically and preserve the spherical nature of the WD.

Spectropolarimetric observations of SNe\,Ia beyond 20 days after maximum light are very rare. The handful of SNe\,Ia for which such measurements exist display low continuum polarization. For example, SN\,2012fr was polarized to $\sim 0.2\%$ on day +24 \citep{Maund_etal_2013}, and SN\,2001el and SN\,2006X showed $<0.1\%$ polarization on days +38 and +39, respectively \citep{Patat_etal_2009, Wang_etal_2003_01el}. These observations challenge the apparent late-time increase in polarization of SN\,2019ein. However, owing to a small sample and the fact that these SNe display a diverse set of properties (expansion velocities, decline rates, etc.), the question remains whether any SNe\,Ia display late-time increases in polarization. More polarimetric observations of SNe\,Ia at similar phases are required to answer that question.

\subsection{Implications for Explosion Scenarios}

The time-invariant $PA$ observed from early times to the phase just prior to 
the secondary maximum in the NIR light curves of SN\,2019ein brings us to the 
implications for explosion scenarios and the characteristics of $p$. Based on 
current understanding, SNe\,Ia might be triggered through the following 
mechanisms.
\begin{enumerate}

\item Deflagration: 
compressional heating in a slow accretion near the WD 
centre triggers subsonic burning when the WD approaches the Chandrasekhar mass 
$M_{\text{Ch}}$ \citep{Nomoto_etal_1984, Gamezo_etal_2004, Ropke_2007, Ma_etal_2014}.

\item Delayed detonation: explosion begins at the end of a deflagration phase near or at the centre of the WD. When a critical density of $\sim 10^7$\,g\,cm$^{-3}$ is reached a transition to detonation occurs \citep{Khokhlov_etal_1991}.

\item Colliding/inspiraling WDs:  heat released on dynamical timescales triggers a detonation of a double-degenerate system \citep{Iben_Tutukov_1984, Webbink_1984,
Benz_etal_1990, Nugent_etal_1997, Pakmor_etal_2010, Kushnir_etal_2013, GB_2017}.

\item Double/helium detonation: a sub-$M_{\text{Ch}}$ C-O WD may explode by
detonating a thin surface He layer, which triggers a detonation front in the 
WD \citep{Woosley_etal_1980, Nomoto_1982_p1, Nomoto_1982_p2, Livne_1990, 
Woosley_Weaver_1994, Hoeflich_Khokhlov_1996, Kromer_etal_2010}.
\end{enumerate} 

In the framework of off-centre 
delayed-detonation models, a slightly asymmetric excitation will lead to an 
off-centre distribution of iron-group elements. The axis of symmetry 
will be defined by the centroid of the density distribution and the point 
of off-centre delayed-detonation transition \citep{Livne_1999, 
Hoeflich_etal_2006, Fesen_etal_2007}. If the $^{56}$Ni region is above 
the photosphere, the time-invariant $PA$ observed in SN\,2019ein would 
indicate a moderate asphericity of the central energy source. Otherwise, a change of $PA$ may be seen as the SN approaches its secondary peak. The low continuum polarization observed in SN\,2019ein challenges any model that predicts significant asymmetry of the photosphere. 

Dynamical or head-on collisions of WDs are expected to show larger 
asymmetries \citep{Benz_etal_1990, Pakmor_etal_2011, Sato_etal_2016, 
Katz_etal_2016, Garcia_Berro_etal_2017, GB_2017}. The dynamical models 
are not favoured since they predict high polarization levels at early 
phases \citep{Pakmor_etal_2012, Bulla_etal_2016a} and larger asymmetries 
in the inner layers or off-centred energy sources, which are 
incompatible with our observations of SN\,2019ein.

%\textbf{\citet{Maund_etal_2013} argued that a low level of continuum polarization ($\lesssim$0.1\%) observed in SN\,2012fr between days $-$11 to $+$24, is inconsistent with the asymmetric distribution of $^{56}Ni$ predicted by the violent merger model of \citet{Pakmor_etal_2012}. Hydrodynamic simulations conducted by \citet{Bulla_etal_2016a} show that at early times, the continuum polarization exhibits a modest viewing angle dependence in violent mergers of C-O WDs (for example, from $\sim$0.3 to 1\% as viewed from the equatorial plane to pole on.) Such simulations of violent mergers of two C-O WDs produce significant asymmetries in both IMEs (e.g., Si and Ca) and $^{56}$Ni. 
%We hereby remark that the potential correlation between the asphericity of $^{56}Ni$ and the continuum polarization level requires quantitative investigation.}

For sub-$M_{\text{Ch}}$ explosions through a double/helium detonation, outer 
asymmetry is expected owing to the He-ignition process. Classical 
He-detonation models require a significant He-surface mass on the order of 
0.01 to 0.1\,$M_{\odot}$ (e.g., \citealp{Nomoto_1982_p1, Nomoto_1982_p2, 
Woosley_Weaver_1994, Hoeflich_Khokhlov_1996, Bildsten_etal_2007, 
Shen_etal_2009}). Starting from the SN explosion, the photosphere recedes 
and will first cross the burning product of the outermost He layer. 
The \ion{O}{i} $\lambda7774$ feature is prominent at day $-$10.9 in the 
flux spectrum. However, we see no polarization signal at the corresponding 
wavelength, suggesting that oxygen was present in the outer layer and 
maintained a spherically symmetric distribution in the expanding envelope. This is in contrast with the $\sim 0.4\%$ \ion{O}{i} $\lambda7774$ line polarization predicted by \citet{Bulla_etal_2016b} for the double/helium detonation models. 
Therefore, we infer that the outermost layer is dominated by the 
spherical pre-explosion C and O from the WD, rather than a He shell, 
since the latter is likely to produce an abundance jump in O in the line-forming region (see, e.g., \citealp{Yang_etal_2020}). In fact, SNe\,2006X and 2004dt, both of which are high-velocity SNe\,Ia, also show no polarization across \ion{O}{i}, putting strong constraints on the distribution of oxygen in the explosion ejecta. We note that the polarimetric properties of more modern models of sub-$M_{\text{Ch}}$ double/helium detonation (e.g., \citealp{Shen_etal_2018}) are currently theoretically unexamined.

\subsection{Comparison with a Detonating Failed Deflagration (DFD) Model}
\citet{Kasen_Plewa_2007} made theoretical predictions for spectropolarimetric 
observations of a $M_{\text{Ch}}$ WD using the detonating failed 
deflagration (DFD) model. They studied one particular model in detail, 
named Y12, in which the WD's ignition starts within a small spherical 
region, 50\,km in diameter and offset 12.5\,km from the centre. 
Here, we compare the observed spectropolarimetric properties of SN\,2019ein with the predictions of \citet{Kasen_Plewa_2007}. 

According to \citet{Kasen_Plewa_2007}, if an SN is observed from the 
deflagration side -- the side where ignition began -- high ejecta velocities are expected. Since SN\,2019ein exhibited very high expansion velocities, we may be observing the explosion from the ignition side (viewing 
angles $\theta \approx 0^{\circ}$). From this orientation, the projected surface 
of the intrinsically ``egg-shaped'' density structure in the observer's 
direction would be fairly circular, leading to low continuum polarization. 
In fact, \citet{Kasen_Plewa_2007} argue that low continuum polarization is 
expected from all viewing angles at peak brightness. Therefore, continuum 
polarization is not informative for constraining the viewing angle of the SN.

The Y12 model predicts substantial line polarization (1--2\%) depending on the viewing angle (Fig. 13 of \citealp{Kasen_Plewa_2007}).
Indeed, we observe significant line polarization across both \ion{Si}{ii} $\lambda$6355 and \ion{Ca}{ii} NIR3 features in SN\,2019ein, which suggests a viewing angle of $\theta \approx 0^{\circ}$ or $90^{\circ}$. A viewing angle of $\sim180^{\circ}$ (opposite to the ignition side) is disfavoured because we observe high polarization across both \ion{Si}{ii} and \ion{Ca}{ii}, whereas in the Y12 model only \ion{Si}{ii} polarization is seen for angles $\sim180^{\circ}$. Together with the high expansion velocity of SN\,2019ein, $\theta \approx 0^{\circ}$ is favoured over other orientations, strengthening the case that we may be viewing the SN from the ignition side. According to \citet{Kasen_Plewa_2007}, such events are rare and expected to constitute roughly $10\%$ of all SNe\,Ia. Spectropolarimetry of more SNe\,Ia is needed to test this prediction.

As described in Section~\ref{quplane}, both \ion{Si}{ii} and \ion{Ca}{ii} display higher degree of clumpiness at early epochs than near and after peak brightness. This is also expected in the DFD model, which can produce a clumpy outer layer of IMEs but maintain a relatively smoother IME distribution in the inner layers.

\subsection{The \ion{Si}{ii} $\lambda$6355 Polarization Compared with a Larger Sample}

We notice that the continuum polarization of SN\,2019ein on the Stokes $q$--$u$ 
diagram can be fitted with straight lines (e.g., see the left panels 
of Fig.~\ref{fig:all_qu}). 
A dominant axis is present in Kast spectropolarimetry between days $\sim -11$ and $+10$. Except for the first epoch, the direction of the dominant axis appears to be unchanged, suggesting that different layers of the ejecta share a roughly fixed axial symmetry. 
These properties indicate that SN\,2019ein belongs to 
the spectropolarimetric type D1 \citep{Wang_Wheeler_2008}, in which a 
dominant axis is identifiable but with significant scatter.

We compared the polarimetric properties of SN\,2019ein with those of
SNe\,2004dt and 2006X, both of which display high expansion velocities 
at early phases. As inferred from the absorption minimum of 
the \ion{Si}{ii} $\lambda$6355 line, SN\,2004dt shows an expansion 
velocity of $\sim 17,000$\,km\,s$^{-1}$ at $\sim 6$--8\,days before 
the optical maximum \citep{Wang_etal_2006_04dt}, and SN\,2006X 
exhibits an expansion velocity of 20,700\,km\,s$^{-1}$ at day 
$-$11.3 \citep{Wang_etal_2008}. A linear correlation between the 
maximum polarization measured across \ion{Si}{ii} $\lambda$6355, 
$p_{\rm Si{\sc II}}^{\rm max}$\footnote{The peak of the 
\ion{Si}{ii}$\lambda$6355 polarization is measured between 
roughly days $-$11 and $+$1 \citep{Cikota_etal_2019}} and the SN 
expansion velocity traced by the same line at day $-$5, 
$v_{\rm Si{\sc II}@-5d}$, has been found by \citet{Cikota_etal_2019} 
based on the analysis of a sample of 35 SNe\,Ia. 
The velocity-polarization relation connects the kinematics with the 
ejecta asymmetry and indicates that a higher departure from spherical 
symmetry for Si is produced at higher velocities.

For comparison with the \ion{Si}{ii} velocity-polarization relation 
presented in Figure~13 of \citet{Cikota_etal_2019}, we estimated 
$v_{\rm Si{\sc II}@-5d} = 15,100 \pm 300$\,km\,s$^{-1}$ for SN\,2019ein. 
The peak polarization of SN\,2019ein across the \ion{Si}{ii} $\lambda$6355 
line, i.e., $p_{\rm Si{\sc II}}^{\rm max}$ derived based on 100\,\AA\ and 
50\,\AA\ binnings on day $-$4, is $0.76 \pm 0.10$\% and 
$0.82 \pm 0.16$\%, respectively. These values place SN\,2019ein 
slightly above the predicted \ion{Si}{ii} $\lambda$6355 polarization. 
We remark that SN\,2019ein is still broadly consistent with the 
\ion{Si}{ii} velocity-polarization relation, corroborating the trend 
that higher-velocity SNe\,Ia tend to exhibit higher polarization. 
SN\,2019ein shows significantly lower \ion{Si}{ii} $\lambda$6355 
polarization compared to SN\,2004dt ($14,870 \pm 140$\,km\,s$^{-1}$, 
$1.34 \pm 0.14$\%), which exhibits an exceptionally high peak 
polarization across the \ion{Si}{ii} line and was considered an 
outlier by \citet{Cikota_etal_2019}. On the other hand, SN\,2006X 
($17,040 \pm 90$\,km\,s$^{-1}$, $0.63 \pm 0.05$\%) is in good agreement 
with the \ion{Si}{ii} velocity-polarization relation.

\citet{Wang_etal_2007} also derived a correlation between the maximum 
line polarization of \ion{Si}{ii} $\lambda$6355 and $\Delta m_{15}(B)$. 
For the former parameter, the observations are often converted to the 
level at five days before the $B$-band maximum, i.e., 
$p_{\rm Si{\sc II}}^{\rm corr-5}$. Owing to the sparsely sampled 
spectropolarimetry, we adopt the peak \ion{Si}{ii} $\lambda$6355 
polarization measured at day $-$4 for SN\,2019ein. The $B$-band 
light-curve decline rate of SN\,2019ein has been determined as 
$\Delta m_{15}(B) = 1.36 \pm 0.02$\,mag \citep{Kawabata_etal_2020} and 
$\Delta m_{15}(B) = 1.40 \pm 0.004$\,mag \citep{Pellegrino_etal_2020}. 
We infer that SN\,2019ein is consistent 
with the $\Delta m_{15}(B)$--$p_{\rm Si{\sc II}}^{\rm corr-5}$ relation 
as presented by \citet{Wang_etal_2007} and \citet{Cikota_etal_2019}. 
This relation can be interpreted such that at a given epoch, higher 
line polarization is expected for less-luminous SNe, which indicates a 
higher chemical nonuniformity. This can be understood as an indication 
that less material is burned in fainter events, and such 
incomplete burning may not be sufficient to wipe out lumpy chemical 
configurations.

Therefore, we conclude that SN\,2019ein is consistent with both the 
\ion{Si}{ii} velocity--polarization relationship and the 
light-curve decline rate--\ion{Si}{ii} polarization relation. Such 
behaviour may be explained with the off-centre delayed-detonation model 
(e.g., \citealp{Hoeflich_etal_2006, Cikota_etal_2019}).

\section{Concluding Summary}
\label{sec:concluding_summary}

We have presented spectropolarimetry of SN\,2019ein, a high-velocity SN\,Ia in 
NGC\,5353. Our observations range from day $-$10.9 to $+$10.1 from the 
$B$-band light-curve peak. We found that the continuum polarization in SN\,2019ein is low, staying $<0.25\%$ until about a month after the explosion. This indicates that the photosphere is quite close to being spherical. 

The blueshifted emission peaks observed in early-time spectra of SN\,2019ein cannot be due to a highly asymmetric explosion, as evidenced by low continuum polarization at early times. However, our observations do not preclude the possibility that optical-depth effects in a steep-density ejecta lead to the apparent blueshift of the emission peaks.

The RINGO3 imaging polarimetry shows an apparent increase in polarization ($\sim 1\%$) around day $+$21. However, owing to significant systematic uncertainties found in previous RINGO3 measurements, we are cautious of the observed rise in polarization. If the post-peak increase in polarization is real and intrinsic to SN\,2019ein, we note that it coincides with the beginning of the transition from \ion{Fe}{iii} to \ion{Fe}{ii} ionisation 
states. The recombination decreases the opacity, providing us a deeper 
view into the SN ejecta. 
We speculate that the possible post-peak rise of the polarization, therefore, could indicate 
the presence of an aspherical central energy source.

The polarization 
position angle does not change drastically over the observed epochs. We also observe high line polarization ($\sim 1\%$) across the \ion{Si}{ii} $\lambda6355$ and the \ion{Ca}{ii} NIR3 features around peak brightness of SN\,2019ein. 
The polarization signatures of SN\,2019ein are consistent with models 
predicting SNe\,Ia explosions that produce a modest amount of asphericity. 
To summarise,
\begin{enumerate}
\item A low amount of asphericity in the high-velocity layers is detected, 
as in previous observations of other SNe\,Ia. 

\item Significant departures from global spherical symmetry can be ruled out throughout the ejecta. A common symmetry axis persists from the outer to the inner layers.

\item After day +21, the possibility of a small amount of polarization caused by an 
asymmetric distribution of $^{56}$Ni, which may arise from many 
different models of SN explosions \citep{Hoeflich_1991, Leonard_etal_2005, Kasen_etal_2009, Pakmor_etal_2010, Seitenzahl_etal_2013, Moll_etal_2014, 
Raskin_etal_2014, Bulla_etal_2015, Bulla_etal_2016a, Bulla_etal_2016b, Hoeflich_etal_2017}, cannot be eliminated.
Spectropolarimetry of more SNe\,Ia at post-peak epochs is needed to confirm whether polarization rises beyond day +20. 

\end{enumerate}

The spectropolarimetric observations of SN\,2019ein 
strengthen existing evidence that the explosions of SNe\,Ia are 
largely spherical, especially when considering that SN\,2019ein is 
an event with one of the highest expansion velocities ever observed.

Finally, we compared the results with the detonating failed deflagration 
model of \citet{Kasen_Plewa_2007} and found that the low continuum 
polarization but high line polarization are consistent with the model. 
A viewing angle of $\theta \approx0^{\circ}$ is favoured, which means we may be viewing SN\,2019ein from the ignition side. 

We recommend high-quality 
spectropolarimetric observations of bright, nearby future SNe\,Ia to be 
carried out covering both their rising and falling phases. Such 
polarimetric tomography is essential for building a robust picture of how 
polarization signatures vary over time and the consequences of various 
explosion mechanisms and progenitor scenarios.

%\vspace{1 cm}
\section*{Acknowledgements}

K.C.P. is thankful to Sergiy Vasylyev and Matthew Chu for helpful discussions.
A.V.F.'s group at U.C. Berkeley acknowledges generous support from the Miller Institute for Basic Research in Science, Sunil Nagaraj, Landon Noll, Gary and Cynthia Bengier, Clark and Sharon Winslow, Sanford Robertson, and many additional donors.
The UCSC team is supported in part by NASA grants NNG17PX03C, 80NSSC19K1386, and 80NSSC20K0953; NSF grant AST-1815935; the Gordon \& Betty Moore Foundation; the Heising-Simons Foundation; and by a fellowship from the David and Lucile Packard Foundation to R.J.F. 
The research of Y.Y. is supported through the Bengier-Winslow-Robertson Fellowship and the Benoziyo Prize Postdoctoral Fellowship. 
P.H. acknowledges the support by the NSF 
project ``Signatures of Type Ia Supernovae, New Physics and Cosmology,'' grant 
AST-1715133. 
The supernova research by L.W. is supported by NSF award AST-1817099. 
J.C.W. is supported by NSF grant AST-1813825. 
The research of J.M. is supported through a Royal Society University Research Fellowship. M.B. acknowledges support from the Swedish Research Council (Reg. No. 2020-03330).

A major upgrade of the Kast spectrograph on the Shane 3\,m telescope at Lick Observatory, led by
Brad Holden, was made possible through generous gifts from the Heising-Simons Foundation, William and Marina Kast, and the
University of California Observatories.
KAIT and its ongoing operation were made possible by donations from Sun Microsystems, Inc., the Hewlett-Packard Company, AutoScope Corporation, Lick Observatory, the U.S. NSF, the University of California, the Sylvia \& Jim Katzman Foundation, and the TABASGO Foundation. 
Research at Lick Observatory is partially supported by a generous gift from Google. We appreciate
the excellent assistance of the staff at Lick Observatory. 

PyRAF, PyFITS, STSCI$\_$PYTHON are products of the Space Telescope Science 
Institute (STScI), which is operated by AURA for NASA.
STScI is operated by the Association of Universities for Research in 
Astronomy, Inc., under NASA contract NAS5-26555. 
This research has made use of NASA's Astrophysics Data System Bibliographic Services;
the SIMBAD database, operated at CDS, Strasbourg, France; and
the NASA/IPAC Extragalactic Database (NED) which 
is operated by the Jet Propulsion Laboratory, California Institute of Technology, 
under contract with NASA.

\section*{Data Availability}
The raw data used in this work may be shared upon request to Kishore C. Patra (kcpatra@berkeley.edu).



\bibliographystyle{mnras}
\bibliography{references} % if your bibtex file is called example.bib

%\begin{thebibliography}{99}
%\end{thebibliography}

%%%%%%%%%%%%%%%%%%%%%%%%%%%%%%%%%%%%%%%%%%%%%%%%%%

%%%%%%%%%%%%%%%%% APPENDICES %%%%%%%%%%%%%%%%%%%%%

%\appendix


%%%%%%%%%%%%%%%%%%%%%%%%%%%%%%%%%%%%%%%%%%%%%%%%%%

% Don't change these lines
\bsp	% typesetting comment
\label{lastpage}
\end{document}